\documentstyle[12pt]{article}

\title{SCALAR--TENSOR THEORIES AND QUANTUM GRAVITY}
\author{FATIMAH SHOJAI\footnote{Email: FATIMAH@NETWARE2.IPM.AC.IR}$^{\ ,1}$
and ALI SHOJAI\footnote{Email: SHOJAI@NETWARE2.IPM.AC.IR}$^{\ ,2,1}$
and MEHDI GOLSHANI\footnote{Email: GOLSHANI@IHCS.AC.IR}$^{\ ,3,1}$\\
$^1$Institute for Studies in Theoretical Physics and Mathematics,
\\P.O.Box 19395--5531, Tehran, IRAN.\\
$^2$Department of Physics, Tarbiat Moddarres University,
\\P.O.Box 14155-4838, Tehran, IRAN.\\
$^3$Department of Physics, Sharif University of Technology,
\\P.O.Box 11365--9161, Tehran, IRAN.}
\date{\today}
\begin{document}
\maketitle
\begin{abstract}
{\it Recently, it was shown that the quantum effects of the matter, could
be used to determine the conformal degree of freedom of the space--time
metric. So both gravity and quantum are geometrical features. Gravity 
determines the causal structure of the space--time, while quantum determines
the scale of the space--time. In this article, it is shown that it is
possible to use the scalar--tensor framework to build a unified theory
in which both quantum and gravitational effects are present.}
\end{abstract}
\section{Quantum Force and Geometry}
As it was discussed in references \cite{Bohm,Holland}, in the 
de-Broglie--Bohm quantum theory, the motion of a hypothetical ensemble 
of quantum particles would be given by the Hamilton--Jacobi equation:
\begin{equation}
g^{\mu\nu}\nabla_\mu S \nabla_\nu S = {\cal M}^2c^2
\end{equation}
in which the mass field is defined as:
\begin{equation}
{\cal M}^2=m^2+\frac{\hbar^2}{c^2}\frac{\Box \sqrt{\rho}}{\sqrt{\rho}}
\end{equation}
and $\rho$ is the ensemble density of the system satisfying the continuity
equation:
\begin{equation}
\nabla_\mu\left ( \rho \nabla^\mu S\right )=0
\end{equation}
The path of the particle could be determined according to the guidance
relation:
\begin{equation}
P^\mu={\cal M}u^\mu=\nabla^\mu S
\end{equation}
It is a simple task to show\cite{QCT} that the Hamilton--Jacobi equation and the guidance
relation lead to the following geodesic equation:
\begin{equation}
\frac{du^\mu}{d\tau}+\Gamma^\mu_{\nu\kappa}u^\nu u^\kappa={\cal F}_Q^\mu
\end{equation}
The right hand side of this equation is the quantum force given by:
\begin{equation}
{\cal F}^\mu_Q=\frac{1}{{\cal M}}\left (g^{\mu\nu}-u^\mu u^\nu \right )
\nabla_\nu {\cal M}
\end{equation}
In reference \cite{QCT}, it was shown that by the conformal transformation:
\begin{equation}
g_{\mu\nu}\longrightarrow\phi g_{\mu\nu}
\label{PHM}
\end{equation}
\begin{equation}
\phi=\frac{{\cal M}}{m}
\end{equation}
the equation of motion would be transformed to an equation in which there
is no quantum
effects. As a result, the geodesic equation would be changed
to the one without the quantum force. This means that it is possible to
have two identical pictures for investigating the quantal effects
of matter in the curved space--time background. According to the first
picture, the space--time metric is $g_{\mu\nu}$ which contains
only the gravitational effects of matter. The quantum effects affect the
path of the particles via the quantum force ${\cal F}_Q^\mu$. In the second
picture, the space--time metric is given by $\phi g_{\mu\nu}$
in which $g_{\mu\nu}$ contains the gravitational and $\phi$ (the 
conformal factor) contains the quantal effects of matter.

This shows that the quantum as well as the gravitational effects
of matter have geometrical nature. The second picture mentioned 
above provides a unified geometrical framework for understanding 
the gravitational 
and quantum forces. Accordingly, we call the metric $\phi g_{\mu\nu}$, the physical
metric (containing both gravity and quantum) and the metric $g_{\mu\nu}$,
the background metric (including only gravity).

In order to have a theory which deals with both gravitational and quantum
aspects of matter, one must write an action principle for such a system.
Before proceeding, it must be pointed out here that the aforementioned
matter can be
interpreted in two ways. According to the first interpretation,  
to bring in quantum effects one needs only to make a scale
transformation (i.e. changing only the metric as is given by (\ref{PHM})). In the second
interpretation one thinks of the relation (\ref{PHM}) as representing a conformal
transformation (i.e.  transformation any physical quantity with an appropriate
power of $\phi$). An action principle appropriate for the first way is written
in \cite{BQG} and for the second way in \cite{QCT}. The equations of motion
in both cases are obtained and the physical properties are investigated.
The important point about both these works is that in order to fix the conformal
degree of freedom of the space--time metric to the one given by the relation 
(\ref{PHM}), the method of lagrange multiplier is used and in this way they are a
little artificial. Here we shall show that in the framework of the scalar--tensor
theories, it is possible to write an action principle, in which both gravitational
and quantum contributions to the geometry are included and that the conformal 
degree of freedom of the space--time metric is fixed at the level of the equations 
of motion not needing the method of lagrange multiplier.
\section{Scalar--Tensor Theories of Gravity}
Scalar--Tensor theories of gravity are purely metric theories describing
the gravitational interactions by the space--time metric and some scalar 
field\cite{Mag}. In the Kalutza--Klein theory, first developed to incorporate both
the gravitational and electromagnetic fields, this scalar field naturally
appears as a component of the five--dimensional metric.

The vaccum lagrangian of the scalar--tensor theory is given by:
\begin{equation}
{\cal L}=\sqrt{-g}\left [ \phi {\cal R}-\frac{\omega(\phi)}{\phi}
\nabla_\mu\phi\nabla^\mu\phi \right ]
\end{equation}
in which $\omega(\phi)$ is a function of the scalar field ($\phi$).
The scalar field is
coupled to the scalar curvature in a non--minimal manner.
This field is an additional degree of freedom of the gravitational
field. So gravity consists of two parts, a spin--2 field (the metric)
and a spin--0 field ($\phi$).
These parts represent the Jordan frame.
The scalar field is affected by matter distribution in the
universe and $\phi^{-1}$ determines the strength of the gravitational
interaction (Satsfication of Mach's principle).
 
One can always make some conformal transformation and redefine the 
fields.
So, the theory can be expressed in terms of an infinite class of 
conformally related frames. One of these transformations is the
Dicke transformation\cite{Dic}, which allows us to change the scalar--tensor 
lagrangian to the Einstein--Hilbert one:
\begin{equation}
{\cal L}=\sqrt{-\tilde{g}}\left [ \tilde{\cal R}-\tilde{\nabla}_\mu\Phi
\tilde{\nabla}^\mu\Phi\right ]
\end{equation}
in which $\tilde{g}_{\mu\nu}=\phi g_{\mu\nu}$, $d\Phi=\sqrt{\omega+3/2}d\phi/\phi$, 
($\omega >-3/2$)and the massless field $\Phi$
is coupled to gravty minimally. These fields, $\tilde{g}_{\mu\nu}$ and $\Phi$,
represent the Einstein's conformal frame. 
Therefore in vaccum any scalar--tensor 
theory is equivallent to general relativity plus a massless scalar field.
The standard minimal interaction between the gravitational matter ($\psi$)
and the space--time metric can distinguish the Jordan and Einstein
conformal frames. In addition, as the scalar field has not yet been 
observed experimentally, one can assume any arbitrary interaction
between this field and the gravitational matter\cite{Sok}. If the ordinary matter is coupled
to gravity minimally in Jordan conformal frame and there is not any
interaction between scalar field and matter, we have:
\begin{equation}
{\cal L}=\sqrt{-g}\left [ \phi{\cal R}-\frac{\omega}{\phi}\nabla_\mu\phi\nabla^\mu\phi
+{\cal L}_m(g,\phi)\right ]
\end{equation}
Now, in the Einstein conformal frame this reads as:
\begin{equation}
{\cal L}'=\sqrt{-\tilde{g}}\left [\tilde{\cal R}-\tilde{\nabla}_\mu\Phi
\tilde{\nabla}^\mu\Phi+\frac{1}{\phi^2(\Phi)}{\cal L}_m(\tilde{g}/\phi(\Phi),\psi)
\right ]
\end{equation}
Thus the scalar field interacts with matter.

Similarly the minimal interaction in Einstein conformal frame leads to
the interaction between scalar field and matter in Jordan conformal frame.
Therefore, by considering the matter coupling, the Einstein and Jordan
conformal frames are two physically different frames. 
Now, the important question is:
{\it "which frame contains the space--time metric of the
physical world?"\/} Many arguments are presented in
the literature for both alternatives\cite{Sok}.

In general, the important
aims in constructing a scalar--tensor theory are the satisfication of Mach's 
principle, the investigation of the effects of the scalar field on the 
early universe singularity, the importance of the scalar field for constructing inflationary
models, and so on. Finally, it must be noted that the scalar--tensor theory
is one of the most important candidates for quantum gravity. This property
is important for us here. 
\section{Quantum Force and Scalar--Tensor Theories}
In the present work we use an appropriate scalar--tensor action along with 
matter action to derive all the equations of quantum gravity. These
equations are the generalized Einstein's equations, the equation of motion
of the conformal factor (which relates this factor and the quantum potential), 
and  the Bohm's equations of motion including the continuity equation and the 
quantum Hamilton--Jacobi equations. 

Now, we start from the most general scalar--tensor action:
\begin{equation}
{\cal A}=\int d^4x \left \{ \phi {\cal R}-\frac{\omega}{\phi}\nabla^\mu\phi
\nabla_\mu\phi+2\Lambda\phi+{\cal L}_m\right \}
\end{equation}
in which $\omega$ is a constant independent of the scalar field, and $\Lambda$
is the so--called cosmological constant. Also, it is
assumed that the matter lagrangian is coupled to the scalar field. This coupling
is also present in the generalized Brans--Dicke\cite{Cho}
 or Kalutza--Klein theories\cite{Dam}.
The wave equation of the scalar field and the generalized Einstein's equations
can be written as:
\begin{equation}
{\cal R}+\frac{2\omega}{\phi}\Box\phi-\frac{\omega}{\phi^2}\nabla^\mu\phi
\nabla_\mu\phi+2\Lambda+\frac{\partial {\cal L}_m}{\partial\phi}=0
\label{PE}
\end{equation}
\begin{equation}
{\cal G}^{\mu\nu}-\Lambda g^{\mu\nu}=-\frac{1}{\phi}{\cal T}^{\mu\nu}
-\frac{1}{\phi}[\nabla^\mu\nabla^\nu-g^{\mu\nu}\Box ]\phi+\frac{\omega}{\phi^2}
\nabla^\mu\phi\nabla^\nu\phi-\frac{1}{2}\frac{\omega}{\phi^2}g^{\mu\nu}
\nabla^\alpha\phi\nabla_\alpha\phi
\end{equation}
where ${\cal G}_{\mu\nu}={\cal R}_{\mu\nu}-1/2{\cal R}g_{\mu\nu}$ is 
Einstein's tensor.

The scalar curvature can be evaluated from the contracted form of the latter
equation, and it can be substituted in the relation (\ref{PE}). Then we have:
\begin{equation}
\frac{2\omega-3}{\phi}\Box\phi=-\frac{{\cal T}}{\phi}+2\Lambda-
\frac{\partial{\cal L}_m}{\partial\phi}
\label{PTE}
\end{equation}
The matter lagrangian for an ensemble of relativistic particles of mass $m$
is (without any quantum contribution): 
\begin{equation}
{\cal L}_{m(no-quantum)}=\frac{\rho}{m}\nabla_\mu S\nabla^\mu S-\rho m
\end{equation}
This lagrangian can be generalized if one assumes that 
there is some interaction between the scalar field and the matter field.
Here, for simplicity, it is assumed that this interaction is in the form of
powers of $\phi$.
In order to bring in the quantum effects, one needs to add terms containig
the quantum potential. 
Physical intuition leads us to the fact that it is
necessary to assume some interaction between cosmological constant and
matter quantum potential. This suggestion will be confirmed after
obtaining all of the equations of motion.
These arguments lead us to consider the matter lagrangian as:
\begin{equation}
{\cal L}_m=\frac{\rho}{m}\phi^a\nabla^\mu S\nabla_\mu S-m\rho\phi^b-\Lambda(1+Q)^c
\end{equation}
in which the $a$, $b$, and $c$ constants are fixed later. Therefore the 
energy--momentum tensor is:
\[ {\cal T}^{\mu\nu}=-\frac{1}{\sqrt{-g}}\frac{\delta}{\delta g_{\mu\nu}}
\int d^4x \sqrt{-g} {\cal L}_m=-\frac{1}{2}g^{\mu\nu}{\cal L}_m+\frac{\rho}{m}
\phi^a\nabla^\mu S\nabla^\nu S -\frac{1}{2}\Lambda cQ(1+Q)^{c-1}g^{\mu\nu}\]
\begin{equation}
-\frac{1}{2}\alpha\Lambda c\nabla_\alpha\sqrt{\rho}\nabla_\beta\left (
\frac{(1+Q)^{c-1}}{\sqrt{\rho}}\right )\left [ g^{\mu\nu}g^{\alpha\beta}
-g^{\alpha\mu}g^{\beta\nu}-g^{\beta\mu}g^{\alpha\nu}\right ]
\end{equation}
Using the matter lagrangian and contracting the above tensor, one can 
calculate the first and third terms in the relation (\ref{PTE}). 
The other equations, the continuity equation and the quantum Hamilton--Jacobi
equation, are expressed respectively as:
\begin{equation}
\nabla^\mu\left ( \rho\phi^a\nabla_\mu S\right )=0
\end{equation}
\begin{equation}
\nabla^\mu S \nabla_\mu S=m^2\phi^{b-a}-\frac{1}{2}\Lambda mc\frac{Q}{\rho\phi^a}
(1+Q)^{c-1}+\frac{1}{2}\Lambda mc\alpha\frac{1}{\sqrt{\rho}\phi^a}
\Box\left ( \frac{(1+Q)^{c-1}}{\sqrt{\rho}}\right )
\end{equation}
To simplify the calculations, with due attention to the
equation (\ref{PTE}), one can choose $\omega$ to be $\frac{3}{2}$. Then a perturbative 
expansion for the scalar field and matter distribution density can be used as:
\begin{equation}
\phi=\phi_0+\alpha\phi_1+\cdots
\end{equation}
\begin{equation}
\sqrt{\rho}=\sqrt{\rho_0}+\alpha\sqrt{\rho_1}\cdots
\end{equation}
In the zeroth order approaximation, the scalar field equation gives:
\begin{equation}
b=a+1;\ \ \ \ \ \ \ \ \phi_0=1
\end{equation}
In the first order approaximation one gets:
\begin{equation}
\alpha\phi_1=\frac{c}{2}(1-a)Q+\frac{a}{2}c\tilde{Q}
\end{equation}
in which:
\begin{equation}
\tilde{Q}=\alpha\frac{\nabla_\mu\sqrt{\rho}\nabla^\mu\sqrt{\rho}}{\rho}
\end{equation}

Since the scalar field is the conformal factor 
of the space--time metric, and because of some arguments\cite{BQG,QCT} show
that this field is a function of matter quantum potential, one might 
choose the constant $a$ equall to zero. Then, the scalar field
is independent of $\tilde{Q}$ and we have:
\begin{equation}
\alpha\phi_1=\frac{c}{2}Q
\end{equation}
Also the Bohmian equations of motion give:
\begin{equation}
\nabla_\mu S\nabla^\mu S=m^2(1+cQ/2)-\Lambda mc\frac{Q-\tilde{Q}}{\rho_0}
\end{equation}
It is necessary to choose $c=2$ in order that the first term on the right
hand side be the same as the quantum mass ${\cal M}$.
These choises for parameters $a$, $b$ and $c$ lead to 
the non--perturbative quantum gravity equations
as follows:
\begin{equation}
\phi=1+Q-\frac{\alpha}{2}\Box Q
\label{YYY}
\end{equation}
\begin{equation}
\nabla^\mu S\nabla_\mu S=m^2\phi-\frac{2\Lambda m}{\rho}(1+Q)(Q-\tilde{Q})
+\frac{\alpha\Lambda m}{\rho}\left ( \Box Q -2\nabla_\mu Q\frac{\nabla^\mu 
\sqrt{\rho}}{\sqrt{\rho}}\right )
\label{ZZZ}
\end{equation}
\begin{equation}
\nabla^\mu(\rho\nabla_\mu S)=0
\end{equation}
\begin{equation}
{\cal G}^{\mu\nu}-\Lambda g^{\mu\nu}=-\frac{1}{\phi}{\cal T}^{\mu\nu}
-\frac{1}{\phi}[\nabla^\mu\nabla^\nu-g^{\mu\nu}\Box ]\phi+\frac{\omega}{\phi^2}
\nabla^\mu\phi\nabla^\nu\phi-\frac{1}{2}\frac{\omega}{\phi^2}g^{\mu\nu}
\nabla^\alpha\phi\nabla_\alpha\phi
\label{TTT}
\end{equation}

We conclude this section by pointing out some important hints:
\begin{itemize}
\item It is very interesting that in the framework of the scalar--tensor
theories, one is able to derive all of the quantum gravity equations of motion
without using the method of lagrange multipliers as it is done in the 
previous works\cite{QCT,BQG}.
\item In the suggested quantum gravity theory, the causal structure of the
space--time ($g_{\mu\nu}$) is determined via equation (\ref{TTT}). This 
shows that except for back--reaction terms of the quantum effects on $g_{\mu\nu}$,
the causal structure of the space--time is determined by the gravitational
effects of matter. Quantum effects, determine directly the scale factor 
of the space--time, from the relation (\ref{YYY}).
\item It must be noted that the mass field given by the right hand side
of the relation (\ref{ZZZ}), consists of two parts. The first part which
is proportionnal to $\alpha$, is a purely quantum effect, and the second part
which is proportional to $\alpha\Lambda$, is a mixture of the quantum effects 
and the large scale structure introduced via the cosmological constant.
\end{itemize}
\section{Concluding Remarks}
This formulation of quantum gravity leads us to the physical frame naturally.
This frame is determined by matter distribution uniqely.
Thus, one may conclude this point is confirmed that the physical frame can not be specified without
matter fields. Therefore the matter distribution determines the local curvature
of the space--time (in confirmity with Mach's principle). Furthuremore,
from the matter equation of motion one can see that the cosmological constant
(a large scale structure constant) and the quantum potential\footnote{In
Bohmian quantum mechanics, observable effects of quantum potential
may appear at both large and small scales, depending on the shape of
the ensemble density\cite{Holland}.} are coupled togheter.
This is another manifestation of the Mach's principle.

The guiding equation $P^\mu={\cal M}u^\mu=\nabla^\mu S$, leads us to
the geodesic equation. From relation (\ref{ZZZ}) we have:
\begin{equation}
\frac{du^\mu}{d\tau}+\Gamma^\mu_{\nu\kappa}u^\nu u^\kappa=\frac{1}{\cal M}
(g^{\mu\nu}-u^\mu u^\nu)\nabla_\nu {\cal M}
\end{equation}
where
\begin{equation}
{\cal M}^2=m^2\phi-\frac{2\Lambda m}{\rho}(1+Q)(Q-\tilde{Q})
+\frac{\alpha\Lambda m}{\rho}\left ( \Box Q -2\nabla_\mu Q\frac{\nabla^\mu 
\sqrt{\rho}}{\sqrt{\rho}}\right )
\end{equation}
Thus, in the present theory, the scalar field produces quantum force that appears
on right hand side and violates the equivalance principle. Similarly, in
Kalutza--Klein theory, the scalar field (dilaton) produces some fifth force 
leading to the violation of the equivalance principle\cite{CP}.


\begin{thebibliography}{99}
\bibitem{Bohm}
D. Bohm, Phys. Rev., {\bf 85}, No. 2, 166, 1952;\\
D. Bohm, Phys. Rev., {\bf 85}, No. 2, 180, 1952;\\
D. Bohm and B.J. Hiley, {\it The undivided universe}, 
Routledge, London, 1993.
\bibitem{Holland}
P.R. Holland, {\it The Quantum Theory of Motion\/},
Cambridge University Press, 1993.
\bibitem{QCT}
F. Shojai, A. Shojai, and M. Golshani, {\it Conformal Transformations
and Quantum Gravity\/}, to appear in Mod. Phys. Lett. A.
\bibitem{BQG}
F. Shojai, and M. Golshani, Int. J. Mod. Phys. A., Vol. 13, No. 4, 677, 1998.
\bibitem{Dic}
R.H. Dicke, Phys. Rev., {\bf 125}, 2163, 1962.
\bibitem{Sok}
L.M. Sokolowski, Talk given at the 14th Conference on General Relativity
and Gravitation, Florence, 1995, {\bf GR--QC/9511073}.
\bibitem{Mag}
G. Magnano, Talk given at the XI Italian Conference on General Relativity
and Gravitation, Trieste, 1994, {\bf GR--QC/9511027}.
\bibitem{Cho}
Y.M. Cho, Phys. Lett. B, {\bf 186}, 38, 1987.
\bibitem{Dam}
T. Damour, G. W. Gibbons, and C. Gundlach, Phys. Rev. Lett., {\bf 64},
123, 1990.
\bibitem{CP}
Y.M. Cho, and D.H. Park, Nuvou Cimento, {\bf 105B}, 817, 1990.
\end{thebibliography}
\end{document}